# Commercially available Geiger mode single-photon avalanche photodiode with a very low afterpulsing probability

Mario Stipčević[1]

[1]Photonics and Quantum Optics Research Unit, Centre of Excellence for Advanced Materials and Sensing Devices, Ruđer Bošković Institute, Bijenička cesta 54, HR-10000 Zagreb, Croatia
*Corresponding author: mario.stipcevic@irb.hr

Date: May 15, 2015

**Abstract.** Afterpulsing is one of the main technological flaws present in photon counting detectors based on solid-state semiconductor avalanche photodiodes operated in Geiger mode. Level of afterpulsing depends mainly on type of the semiconductor, doping concentrations and temperature and presents an additional source of noise, along with dark counts. Unlike dark counts which appear randomly in time, aterpulses and are time-correlated with the previous detections. For measurements that rely on timing information afterpulsing can create fake signals and diminish the sensitivity. In this work we test a novel broadband sensitive APD that was designed for sub-Geiger avalanche gain operation. We find that this APD, which has a reach-through geometry typical of single-photon detection photodiodes, can also operate in Geiger mode with usable detection sensitivity and acceptable dark counts level while exhibiting uniquely low afterpulsing. The afterpulsing of tested samples was systematically less than 0.05 percent at 10V excess voltage.

Detectors of extremely low levels of light (photon counters) based on semiconductor avalanche photodiodes are increasingly used in many areas of science, industrial and health applications, such as: quantum information [1], quantum cryptography [2], quantum random number generators [3-5], fluorescence correlation and single molecule spectroscopy (FCS) [6], low-light imaging [7] and holography [8], particle sizers [9], gas flow monitors [10], medical diagnostic equipment [11] etc. The technology of semiconductor silicon single photon capable avalanche diodes (SPADs) started in 1960's [12] and is ever since a very active area of research. Silicon detectors are sensitive to a broad spectrum of wavelengths including the whole visible spectrum of light and achieve a detection time resolution (jitter), quantum efficiency and gain that significantly surpass those of traditional photomultiplier tubes. In addition to the robust build, small size and low power consumption, they can be incorporated directly into a silicon chip using existing chip-making technology, which opens the possibility of innumerable applications. most notably photon counting cameras that started to emerge in past few years [7], [13-14].

All internal-gain single photon sensors suffer, to a greater or lesser extent, from the same family of imperfections, notably from: non-unity detection efficiency, dark counts, afterpulsing, detection time jitter.

In this study we are primarily interested in afterpulsing probability which we define as probability that a first detection event following a known photon detection is not a dark count.

Afterpulsing is one of the main technical flaws present in photon counting detectors based on solid-state semiconductor avalanche photodiodes operated in Geiger mode. Afterpulsing is due to deep level states that are filled by charge created by an avalanche and that decay after a characteristic lifetime(s). Level of afterpulsing and lifetime depends mainly on type of the semiconductor, concentration of point defects and temperature. Silicon SPADs suffer from non-negligible aterpulsing probability, typically in the range of 1-10% with lifetime(s) in range ~10 ns to ~1 µs, while InGaAs(P) exhibit about 5-15% and have about an order of magnitude longer lifetimes.

Both dark counts and afterpulses create fake detection signals that are physically undistinguishable from light photon detections, but unlike dark counts which appear randomly in time, aterpulses and are time-correlated with the previous detections. For measurements that rely on timing information, such as time-resolved spectroscopy, afterpulsing directly generates a fake signal and diminishes the sensitivity by elevating the noise floor.

In this work we test a novel broadband sensitive APD, SUR500 from Laser Components, that was designed for sub-Geiger avalanche gain operation.

We find that this APD, which has a reach-through geometry [15] typical of single-photon detection photodiodes, can also operate in Geiger mode with usable detection efficiency, dark counts rate and jitter, while at the same time exhibiting, to best of our knowledge, a uniquely low afterpulsing probability.

**Experimental setup**

Experimental setup consists of the detector which is coupled to a picosecond pulsed laser (pulse FWHM 39ps, wavelength 676 nm) via a mode-mismatching adjustable coupler.

The detector has been build according to Ref. [16] witn an improved quenching step of 30 V that enabled active quenching at overvoltages as high as 25 V.

Two methods of determining afterpulses were employed. First, the laser was switched off leaving detector in darkness. Using time-to-amplitude converter (TAC) waiting times between subsequent pairs of dark count events were recorded to a PC computer and histogrammed in 100 bins thus obtaining a representation of probability density function (p.d.f) of the waiting times. Since physical dark events are exponentially distributed, any excess events above the exponential in the short end of the probability density distribution will be interpreted as afterpulsing [17-19]. For a Si APD operated at -20 °C one expects main afterpulsing occurring within a few hundreds of nanoseconds with most pronounced lifetime on the order 20-40 ns as found in SAP500 [18] and SLiK [20] reach-through AP diodes. Consequently an initial TAC range od 1 μs was chosen for recording afterpulses and dark counts. As no afterpulsing was observed, the search for afterpulsing was extended towards longer times. In all, data was independently collected in four TAC ranges: 1 μs, 5 μs, 20 μs and 200 μs. The statistics was about 100,000 intervals in each range. Obtained distributions are shown in semilog scale in Fig. 1 (a)-(d). The hypothesis that the empirical distribution is consistent with a cut-off exponential p.d.f. (cut at the low end due to the dead time of the detector and at the high end due to the finite range of the TAC) was statistically tested by $\chi^2$ test. All four distributions conform with the exponential hypothesis within 95% CL.

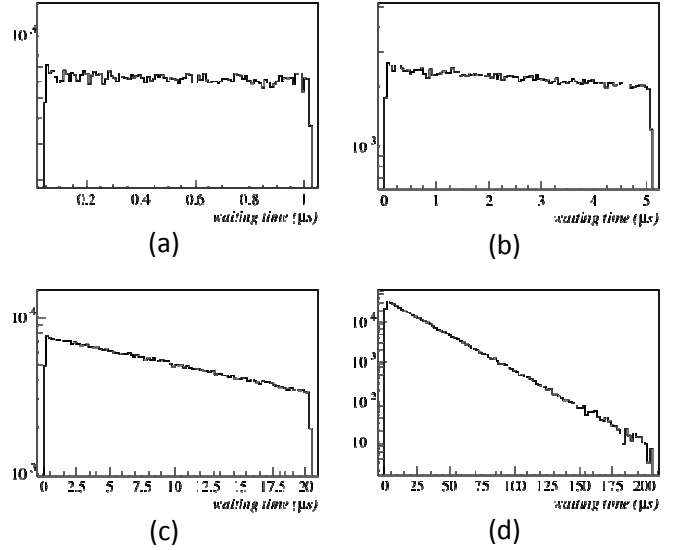

Fig. 1. Waiting time distributions of dark counts of SUR500 obtained in Geiger mode at operating temperature of -20 deg °C and 15 V excess voltage.

In order to set tighter limits on afterpulsing probability, a more precise method based on hardware developed in Ref. [18] was used. The principle of this method is the following. A single photon is sent to the detector and if it gets detected then the next detection occurring during a predefined time period $\Delta T$ is counted as a "secondary" event. Since no afterpulses were apparent in the previous test and most probable lifetime for Si reach-through APD is on the order of a few tens of nanoseconds, we used $\Delta T = 500$ ns which should be enough to collect virtually all afterpulses. Let the ratio of number of detected secondary events to the number of detected photons be denoted by $r$, then the afterpulsing probability $p_a$ is given by [18]:

$$p_a = r - \Delta T \cdot f_{DC} \qquad (1)$$

where $f_{DC}$ is the dark count rate. The result for thermoelectrically cooled sample in TO-8 case is $p_a$= 0.0002 or 0.02%. To best of our knowledge this is by far the smallest reported afterpulsing probability for a Si APD in Geiger mode.

**Conclusion**

We tested 5 samples of new silicon avalanche photodiode SUR500 and found that distinguishing characteristics of this APD is its very low afterpulsing probability that was find to be less than 0.05 percent at 15V excess voltage for the tested pool of samples.

Furthermore, all can be consistently operated in Geiger mode serving as a sensitive single-photon detector.

The pool of samples has shown relatively small and near-Gaussian dispersion of Geiger breakdown voltage and dark count rates which all indicates consistent manufacturing.

**Acknowledgment**

This work was supported by the Croatian Ministry of Science Education and Sports, contracts 098-0352851-2873 and 533-19-14-0002.